\documentclass[twocolumn,aps,prl,showpacs]{revtex4}
\usepackage{amsmath}
\usepackage{amssymb}
\usepackage{color}
\usepackage[dvips]{graphicx}

\begin{document}

\title{Tunable control of the bandwidth and frequency correlations of entangled photons}

\author{M. Hendrych$^1$, M. Mi\v{c}uda$^{1,2}\footnote{Permanent address: Department of Optics, Palack\'{y}
University, Olomouc, Czech Republic.}$ and J. P. Torres$^{1,2}$ }

\affiliation{ICFO-Institut de Ci\`{e}ncies Fot\`{o}niques$^{1}$,
and Department of Signal Theory and Communications$^{2}$,
Universitat Polit\`{e}cnica de Catalunya, Castelldefels, 08860
Barcelona, Spain}

\email{juan.perez@icfo.es}

\begin{abstract}
We demonstrate experimentally a new technique to control the
bandwidth and the type of frequency correlations (correlation,
anticorrelation, and even uncorrelation) of entangled photons
generated by spontaneous parametric downconversion. The method is
based on the control of the group velocities of the interacting
waves. This technique can be applied in any nonlinear medium and
frequency band of interest. It is also demonstrated that this
technique helps enhance the quality of polarization
entanglement even when femtosecond pulses are used as a pump.
\end{abstract}

\pacs{03.67.Mn, 42.50.Dv, 42.65.Lm}
\maketitle

One of the goals of quantum optics is to design and implement new
sources of quantum light which could enable tunable control of the
relevant photonic properties, as required by the specific quantum
information applications under consideration.

To date most quantum information applications use the
polarization of photons, or polarization entanglement between
paired photons, as the quantum resource. In this Letter we
present a new technique to appropriately engineer the
frequency properties of quantum light, namely, the bandwidth and
the type of frequency correlations between paired photons.

The spectrum of photons can be considered as a quantum resource by
itself. The frequency content of light, and frequency
entanglement, occurs in an infinite dimensional Hilbert space.
This offers the possibility to implement quantum algorithms that
either inherently live in a higher dimensional Hilbert space
(qudits) or exhibit enhanced efficiency in increasingly higher
dimensions.

On the other hand, the frequency properties of entangled
two-photon states cannot be neglected even when entanglement takes
place in the polarization or spatial degrees of freedom. Since the
corresponding entangled states make use of only a portion of the
total two-photon quantum state, it is required to suppress
any frequency "which-path" information that otherwise degrades the
degree of entanglement.

The optimum bandwidth as well as the most appropriate type of
frequency correlations between paired photons depend on the
specific quantum information application under consideration.
Atom-photon interfaces address specific atomic transitions that
require ultra-narrow-band quantum light ($\sim$ MHz), while the
generation of ultrahigh fluxes of entangled photons ($\sim \mu $W)
while maintaining their nonclassical properties requires light
with largely enhanced bandwidth ($\sim$ tens of
nm)~\cite{loudon1,dayan1}. Some protocols for quantum enhanced
clock synchronization and position measurement rely on the use of
frequency correlated photons~\cite{lloyd1}. Heralded single
photons with a high degree of quantum purity can be obtained by
generation of uncorrelated paired photons. The tolerance against
the effects of mode mismatch in linear optical circuits, the
dominant cause of photon distinguishability, can be enhanced by
using photons with appropriately tailored wave-packet
shape~\cite{rohde1}.

The most widely used method for the generation of pairs of
entangled photons is spontaneous parametric downconversion
(SPDC) where two lower-frequency photons are generated when a
strong pump field interacts with a nonlinear crystal. SPDC with
continuous-wave pumping produces frequency anticorrelated
photons. Other specific frequency correlations, such as correlated
or uncorrelated photons, can occur in special crystals with
suitable pump light conditions, specific values of the nonlinear
crystal length and dispersive properties of the nonlinear
crystals~\cite{grice1,kocuzu1}.

One strategy for engineering the bandwidth is based on the proper
preparation of the downconverting crystal. Frequency-entangled
pairs of photons with a largely enhanced bandwidth can be
generated in chirped quasi-phase-matched nonlinear
crystals~\cite{carrasco3}, while paired photons with a largely
reduced spectral width can be generated in cavity SPDC~\cite{ou1}.
Properly designed SPDC configurations with nonlinear crystal superlattices
allow tailoring the frequency correlations of paired photons~\cite{uren1}.

In general, noncollinear SPDC geometries enable to control the
bandwidth of downconverted photons as well as the
waveform~\cite{carrasco2}. SPDC configurations where the downconverted
photons counter propagate allow to reduce the spectral
width of light~\cite{booth1,rossi1}. Noncollinear SPDC allows the
generation of frequency-correlated and uncorrelated photons by
controlling the pump beam width and the angle of emission of the
downconverted photons~\cite{walton1,uren2}.

In this Letter we experimentally demonstrate a new technique to
tailor the bandwidth and frequency correlations of entangled
photons~\cite{torres1} that makes possible to generate
frequency-correlated, anticorrelated, and even uncorrelated
entangled photon pairs. The method is based on the proper
tailoring of the group velocities of all interacting waves
through the use of beams with angular dispersion, i.e., pulses
with pulse-front tilt. The method can be implemented in materials
and frequency bands where conventional solutions do not hold.

Let us consider a collinear SPDC configuration with type-II ({\em
oee}) phase matching. The input pump beam with central angular
frequency $\omega_p$ passes through a medium with angular
dispersion, such as a diffraction grating oriented along the transversal
$x$-direction. The beam is diffracted by the grating so that each
frequency component is dispersed in a different direction. The
transformation of the pump beam due to the grating can be written
as~\cite{martinez1} $E_p \left(\omega_p+\Omega_p,p_x,p_y \right)
\Longrightarrow E_p \left(\omega_p+
\Omega_p,p_x/\alpha-\Omega_p\,\tan \phi/(\alpha c) ,p_y \right)$,
where $\Omega_p$ is the angular frequency deviation, ${\bf
p}=\left( p_x,p_y \right)$ is the transverse wave-vector,
$\alpha=-\cos \theta_0/\cos \beta_0$, $\theta_0$ is the angle of
incidence at the grating, $\beta_0$ is the output diffraction
angle, and $c$ is the speed of light.

The resulting beam acquires a pulse-front tilt such that its peak
intensity is located at a different time for each value of $x$. The
tilt angle is $\tan \phi=-\lambda_p \epsilon$, where
$\epsilon=m/\left( d \cos \beta_0 \right)$ is the angular
dispersion with $d$
being the groove spacing of the grating, $m$ the diffraction order
and $\lambda_p=2\pi c/\omega_p$. At the output face of the
nonlinear crystal ($z=L$), a second grating is used to
re-collimate the beam by compensating for the angular dispersion
introduced by the first grating.

The quantum state of the downconverted photons at the output face
of the nonlinear crystal writes $|\Psi>=\int d\Omega_s \Omega_i
d{\bf p} d{\bf q} \,\Phi \left( \Omega_s, \Omega_i,{\bf p},{\bf q}
\right) |\omega_s+\Omega_s,{\bf p}>_s\,|\omega_i+ \Omega_i,{\bf
q}>_i$, where~\cite{torres1}
\begin{eqnarray}
\label{state1}
    & & \Psi \left( \Omega_s,\Omega_i,{\bf p},{\bf q} \right)=
    E_p \left( \Omega_s+\Omega_i,{\bar p}_x+{\bar q}_x,p_y+q_y \right)  \nonumber \\
    & & \times sinc
    \left( \Delta_k L/2 \right) \exp \left( i s_k L/2 \right)
    {\cal F} \left( \Omega_s \right) {\cal F} \left( \Omega_i \right)
\end{eqnarray}
and $\Delta_k=k_p \left( \Omega_s+\Omega_i,{\bar
p}_x+\bar{q}_x,p_y+q_y \right)-k_s \left( \Omega_s,\bar{p}_x,p_y
\right)-k_i \left( \Omega_i,\bar{q}_x,q_y \right)$, $s_k=k_p
\left( \Omega_s+\Omega_i,{\bar p}_x+\bar{q}_x,p_y+q_y \right)+k_s
\left( \Omega_s,\bar{p}_x,p_y \right)+k_i \left(
\Omega_i,\bar{q}_x,q_y \right)$, ${\bar p_x}=\alpha
p_x+\Omega_s\,\tan \phi /c$, ${\bar q_x}=\alpha q_x+\Omega_i\,\tan
\phi /c$, and ${\cal F}$ are the corresponding interference
filters in front of the detectors.

Eq. (\ref{state1}) describes an entangled state in both, the
signal and idler transverse wavenumber $({\bf p}, {\bf q})$ and signal and idler
frequency $(\Omega_s, \Omega_i)$. However, in a specific detection
scheme, it has to be projected into the required spatial modes.
Here, the detection scheme projects the signal and idler photons
into two spatial modes with large areas, i.e. ${\bf p}={\bf q}=0$,
so that the quantity of interest is $\Psi \left( \Omega_s,\Omega_i
\right) \equiv \Psi \left( \Omega_s,{\bf p}=0, \Omega_i,{\bf q}=0
\right)$.

The effective inverse group velocity ($u_j$, $j=s,i,p$) of all interacting
waves is~\cite{torres2} $u_j=N_j-\tan \rho_j \tan \phi/c$, and the
effective inverse group velocity dispersion is
$g_j=D_j-\left[ \tan \phi/c \right]^2/k_j$, where $\rho_j$ is the
Poynting vector walk-off, $k_j$ is the longitudinal wavenumber,
and $N_j$ and $D_j$ are the corresponding inverse group velocity
and group velocity dispersion parameters, respectively.

The joint spectral amplitude $\Phi \left( \Omega_s, \Omega_i
\right)$ is determined by the spectral shape of the pump beam and
by the corresponding effective phase-matching conditions inside
the nonlinear crystal. These depend on the relationship between all
the group velocities~\cite{grice1}. For equal effective group
velocities ($u_s=u_i$), one obtains photons highly anticorrelated
in frequency ($\Omega_s=-\Omega_i$). For $u_p=(u_s+u_i)/2$, the
pair of photons is highly correlated in frequency
($\Omega_s=\Omega_i$).

\begin{figure}
\centering\includegraphics[width=0.8\columnwidth]{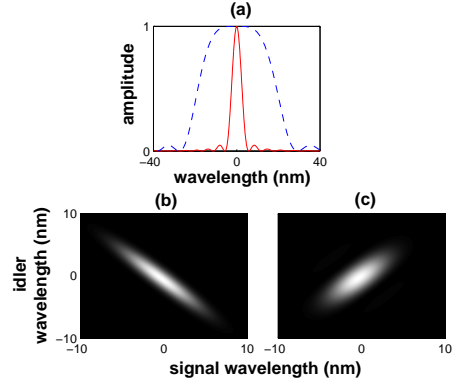}
\caption{Control of the bandwidth and frequency correlations of
paired photons with a tilted pump pulse. (a) Bandwidth of the
signal photons with CW pump for $\phi=0^o$ (solid line) and $\phi=38.1^o$
(dashed line). (b) and (c) Joint spectral intensity for a
broadband pump beam (b) high frequency
anticorrelation ($\phi=38.1^o$) and (c) high frequency correlation
($\phi=-51.9^o$).}
\end{figure}

Let us first consider a continuous-wave (CW) pump beam with
pulse-front tilt. Fig.~1(a) shows how the bandwidth of the joint
spectral intensity $S \left(\Omega_s \right)$ of the downconverted
photons can be tailored by modifying the tilt angle. For
no tilt ($\phi=0$), the spectral intensity writes $S
\left(\Omega_s \right) \propto \rm{sinc}^2 \left[ \left( N_s-N_i
\right)\Omega_s L/2\right]$ which is typical of type-II SPDC
\cite{rubin1}.

The condition $u_s=u_i$ is achieved for a tilt angle \cite{torres1}
\begin{equation}
\label{anticorr}
 \phi=\tan^{-1} \left\{ \frac{c \left( N_i-N_s \right)}{\tan
\rho_s-\tan \rho_i} \right\}
\end{equation}
so that one obtains $S \left(\Omega_s \right) \propto \rm{sinc}^2
\left[ \left( g_s+g_i \right)\Omega_s^2 L/4\right]$, which is
characteristic of a typical type-I SPDC source, although the
process considered is of type-II. We can see that applying a tilt,
the spectrum can vary greatly.

Tilted pulses can also be used to tailor the type of frequency
correlations of photons when broadband pump pulses are used. Let
us define two spectral intensities in the directions
$\Omega_{\pm}=\Omega_s \pm \Omega_i$, $S_{+} \left( \Omega_+
\right)=\int d\Omega_{-} |\Psi \left( \Omega_{+},\Omega_{-}
\right)|^2$ and $S_{-} \left(\Omega_{-} \right)=\int d\Omega_{+}
|\Psi \left( \Omega_{+},\Omega_{-} \right)|^2$, respectively.
Frequency anticorrelated (correlated) photons correspond to the
case where the bandwidth of $S_{+}$ is much narrower (wider) than
the bandwidth of $S_{-}$.

For the tilt angle given by Eq. (\ref{anticorr}), the spectral
intensities $S_{+}$ and $S_{-}$ write
\begin{eqnarray}
\label{anticorr2} & & S_{+} \left( \Omega_+ \right) =|E_p \left(
\Omega_{+} \right)|^2 {\rm sinc}^2 \left( D_+ L\,\Omega_{+}^2/2
\right) {\cal F} \left( \Omega_+ \right), \nonumber \\
& & S_{-} \left( \Omega_{-} \right)= {\rm sinc}^2 \left[ \left( g_s+g_i
\right) L\,\Omega_{-}^2 /16 \right] {\cal F} \left( \Omega_{-}
\right),
\end{eqnarray}
where $D_+=u_p-(u_s+u_i)/2$. The corresponding joint spectral
intensity plotted in Fig.~1(b) shows high frequency
anticorrelation. The phase-matching conditions induced by the
angular dispersion widen the joint spectral intensity in the
$\Omega_{-}$ direction and they make it narrower
in the $\Omega_{+}$ direction.

The tilt required for the generation of highly
frequency-correlated photons, as shown in Fig.~1(c), writes
\begin{equation}
\label{corr}
 \phi=\tan^{-1} \left\{ \frac{c \left( 2N_p-N_s-N_i \right)}{
\tan \rho_s+\tan \rho_i-2\tan \rho_p} \right\}
\end{equation}
and the corresponding spectral intensities are
\begin{eqnarray}
\label{corr2}
& & S_{+} \left( \Omega_+ \right) =|E_p \left( \Omega_{+} \right)|^2 {\cal F} \left( \Omega_+ \right), \nonumber \\
& & S_{-} \left( \Omega_{-} \right)= {\rm sinc}^2 \left[ \left( u_s-u_i
\right) L\,\Omega_{-}/4 \right] {\cal F} \left( \Omega_{-} \right).
\end{eqnarray}
The bandwidth in the $\Omega_{+}$ direction is mostly determined
by the bandwidth of the pump beam, while in the $\Omega_{-}$
direction we obtain the typical type-II dependence on the
frequency.

\begin{figure}
\centering\includegraphics[width=0.8\columnwidth]{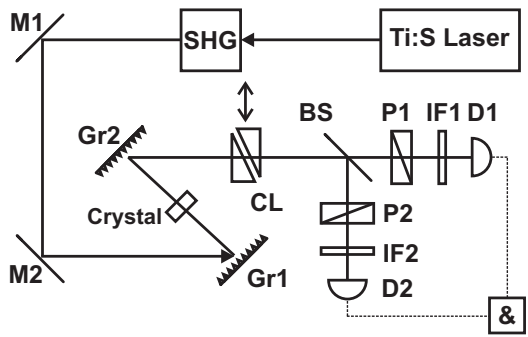}
\caption{Experimental set-up. $SHG$: second harmonic generation;
$M$: mirrors; $Gr$: gratings; $CL$: compensation line; $BS$:
beamsplitter; $P$: polarizers; $IF$: 10-nm interference filters;
$D$: single photon counting modules; \&: coincidence electronics.}
\end{figure}

We set up an experiment (see Fig.~2) to demonstrate the
feasibility of the technique described above. The second harmonic
beam of a femtosecond Ti:Sapphire laser tuned at $810$ nm is
directed at a diffraction grating ($Gr1$) that introduces the
appropriate amount of pulse-front tilt ($\phi$) in the plane
determined by the pump beam direction of propagation and the optic
axis of the nonlinear crystal. The measured bandwidth of the pump
beam was $\Delta \lambda_p=3.6$ nm (FWHM). The pulses diffracted
off the grating enter a 2-mm thick BBO crystal where degenerate
collinear type-II SPDC occurs. After the downconversion process,
the angular dispersion of the downconverted photons is removed by
an inverse grating ($Gr2$).

To evaluate the different types of frequency correlations and the
bandwidth, a Hong-Ou-Mandel (HOM) interferometer is used \cite{hom1}.
A variable polarization-dependent delay line made of birefringent
quartz is inserted between the nonlinear crystal and the
beam splitter, which allows us to add a variable time delay
($\tau$) between the generated paired photons. After the beam
splitter, polarizers ($P$) are used to control the polarization of
the photons. $10$-nm (FWHM) interference filters are located in
front of single-photon counting modules, where the coincident
arrival of two photons in a time window of $\simeq 12$ ns is
measured.

The coincidence rate $R \left(\tau,\theta_a=-45^o,\theta_b=45^o
\right)$, where $\theta_{a,b}$ are the angles of the two
polarizers, is given by
\begin{equation}
\label{hom}
 R \left( \tau \right)=\frac{1}{2} \left[ 1-\frac{1}{2} \int d\Omega_{-} S_0 \left( \Omega_{-} \right)
 \exp \left( -i \Omega_{-} \tau \right) \right],
\end{equation}
where $S_0 \left( \Omega_{-} \right)=\int d\Omega_{+} \Psi \left(
\Omega_{+},\Omega_{-} \right) \Psi^{*} \left(
\Omega_{+},-\Omega_{-} \right)$.

\begin{figure}
\centering\includegraphics[width=0.9\columnwidth]{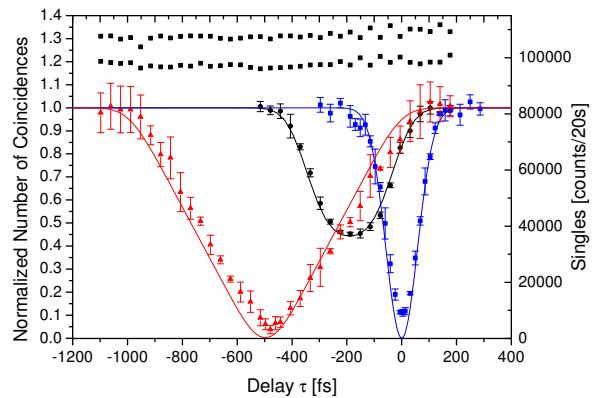}
\caption{Normalized number of coincidences and singles as a
function of time delay in a Hong-Ou-Mandel interferometer for
three different values of the pulse-front tilt of the pump.
Circles: no pulse-front tilt ($\phi=0^o$); squares: highly
frequency-anticorrelated photons ($\phi=38.1^o$); triangles:
highly frequency-correlated photons ($\phi=-51.9^o$); black
squares: singles counts. Solid lines are the theoretical
prediction. The experimental points are raw data without any
corrections for measurement noise.}
\end{figure}

Fig.~3 shows our main experimental results. The normalized number
of coincidences (for the sake of comparison) is plotted versus the
time delay introduced by the delay line. When no tilt is applied
to the pump pulse, the visibility
$V=(R_{max}-R_{min})/(R_{max}+R_{min})$ of the HOM dip drops to
$V=38\%$. The degradation of frequency anticorrelation due to the
broadband pump beam introduces distinguishing information between
the ordinary and extraordinary downconverted photons. The center
of the dip is located at $\tau_0= \left( N_s-N_i
\right)L/2$ \cite{rubin1}.

The tilt angle for anticorrelated photons is $\phi=38.1^o$ and
we expect $\tau_0=0$. This amount of angular dispersion is
introduced by a grating of 1200 lines/mm and diffraction order
$m=1$. Grating $Gr2$ then has 600 lines/mm and $m=-1$. For
anticorrelated photons we obtain $S_0 \left( \Omega_{-}
\right)=S_{-} \left( \Omega_{-} \right)$. The measured visibility
of the HOM dip is now $83\%$. Notice that we generate highly
anticorrelated photons even when we are using a broadband pump
beam.

The generation of highly correlated photons requires a tilt angle
$\phi=-51.9^o$. We expect $\tau_0= \left( u_s-u_i \right) L/2$. A
grating $Gr1$ with 2400 lines/mm and $m=-1$, and $Gr2$ with 1200
lines/mm and $m=1$ are used. The spectral intensity writes $S_0
\left( \Omega_{-} \right)=S_{-} \left( \Omega_{-} \right) \exp
\left[ i \left(u_s-u_i \right)L/2 \right]$.  The visibility of the
HOM dip goes up to $93\%$. The shape of the dip is triangular, as
corresponds to a typical type-II-like SPDC process
\cite{rubin1}. The values of $\tau_0$ measured agree well with the
theoretical predictions. In general, the degree of frequency
correlation or anticorrelation might be modified by using
crystals of different length or filters with different bandwidth.

The use of frequency-correlated (anticorrelated) photons allows the
erasing of the distinguishing information coming from the spectra of
the photons when considering polarization entanglement. Under our
experimental conditions, the two-photon quantum state in
polarization space writes
$|\Psi>=\varepsilon|\psi><\psi|+(1-\varepsilon)/2 \left\{
|H>_s|V>_i<H|_s <V|_i+ |V>_s|H>_i<V|_s <H|_i \right\}$, where
$|\psi>=1/\sqrt{2} \left\{ |H>_s|V>_i+ \exp \left(i\Delta
\right)|V>_s|H>_i \right\}$. The purity of the state is
$P=\left(1+\varepsilon^2\right)/2$ and the visibility of $R
\left(\tau_0,\theta_a=-45^o,\theta_b\right)$ is $V=\varepsilon |\cos \Delta|$.

Fig.~4(a) shows $R \left(\tau_0,\theta_a=-45^o,\theta_b \right)$
for frequency anticorrelated photons when the angle $\theta_b$ is
changed. The corresponding case with no tilt is also shown for
comparison. The visibility increases from $58\%$ (no tilt) to
$88\%$. Fig.~4(b) shows the case for correlated photons with
$\theta_a=-45^o$ and $-30^o$. The visibility goes up to $95\%$ in
both cases. Thus the purity of the state is greater than 0.95.

\begin{figure}
\centering\includegraphics[width=1\columnwidth]{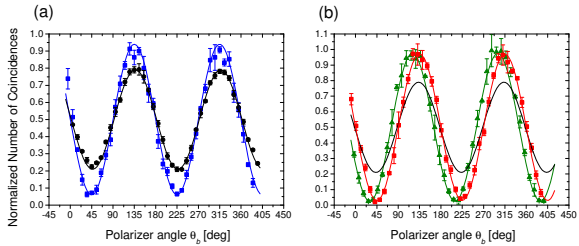}
\caption{Coincidence rate as a function of the polarizer angle
$\theta_b$. (a) Squares: highly anticorrelated photons. Circles:
no tilt. (b) Squares: highly correlated photons with $\theta_a=-45^o$
and triangles with $\theta_a=-30^o$; black line: no tilt.
Solid lines are $\sin^2$-like fits to the raw experimental data.}
\end{figure}

In conclusion, we have shown experimentally that the use of pulses
with pulse-front tilt allows tailoring the bandwidth and frequency
correlations of entangled photons. This technique can be used in
any nonlinear crystal and frequency band of interest. Pulses with
pulse-front tilt are an important and enabling tool in nonlinear
optics, such as in broadband frequency conversion
\cite{martinez2,dubietis1} and soliton phenomena in second order
nonlinear media \cite{paolo1,liu1}. In this Letter, we add the use
of tilted pulses to the toolkit of available techniques in quantum
optics for the full control of the properties of quantum light.

Pulse-front techniques can also be used in other configurations.
In noncollinear SPDC the group velocity can be tailored too, now with
the noncollinear emission angle playing the role of the Poynting-vector
walk-off angle. Therefore, pulse-front techniques can take
advantage of the larger nonlinear coefficient of some materials
in combination with quasi-phase-matching techniques.

The recently demonstrated generation of paired photons in
electromagnetically induced transparency schemes~\cite{balic1} can
also benefit from the use of tilted pulses to enhance the control
of the frequency correlations of the paired photons, as well as to
increase the tuning range of the coherence time of the generated
photons.

This work was supported by projects FIS2004-03556 and
Consolider-Ingenio 2010 QOIT from Spain, and by the European
Commission under the Integrated Project Qubit Applications
(Contract No. 015848). MH and JPT acknowledges support from the
Gov. of Catalonia. MM acknowledges support from the project Center
of Modern optics (LCO6007) of the Czech M. of Education.

\end{document}